\documentclass[conference]{IEEEtran}

\ifCLASSINFOpdf

\else

\usepackage[dvips]{graphicx}

\fi

\usepackage[cmex10]{amsmath}

\usepackage{amsmath,amsthm,amsfonts,amssymb,latexsym,extarrows,booktabs,array}

\begin{document}

\title{Dynamic Power Splitting Policies for AF Relay Networks with Wireless Energy Harvesting}

\author{\IEEEauthorblockN{Lansheng Hu \IEEEauthorrefmark{1}, Chao Zhang \IEEEauthorrefmark{1}\IEEEauthorrefmark{2}}
\IEEEauthorblockA{
\IEEEauthorrefmark{1}School of Electronics and Information Engineering\\
Xi'an Jiaotong University, Xi'an, China\\
E-mail:cjhls8988@stu.xjtu.edu.cn, chaozhang@mail.xjtu.edu.cn\\
\IEEEauthorrefmark{2}National Mobile Communications Research Laboratory,
Southeast University.
}
\and
\IEEEauthorblockN{Zhiguo Ding}
\IEEEauthorblockA{School of Computing and Communications\\
Lancaster University\\
United Kingdom \\
E-mail: z.ding@lancaster.ac.uk}
}

\maketitle

\begin{abstract}
Wireless energy harvesting (WEH) provides an exciting way to supply energy for relay nodes to forward information for the source-destination pairs. In this paper, we investigate the problem on how the relay node dynamically adjusts the power splitting ratio of information transmission (IT) and energy harvesting (EH) in order to achieve the optimal outage performance. According to the knowledge of channel state information (CSI) at the relay, optimal dynamic power splitting policy with full CSI and partial CSI are both provided. Finally, through simulations, the proposed power splitting policies can improve the outage performances and the policy with full CSI achieves the best performance. It is also shown that the policy with partial CSI can approach the policy with full CSI closely and incurs far less system overhead.
\end{abstract}

\IEEEpeerreviewmaketitle

\section{Introduction} \label{sec:1}
Relay-assisted communication is an effective way to improve the transmission reliability and extend the coverage. However, in energy-constrained relay networks, the relay nodes are usually equipped with limited energy supply, such as battery. In many cases, it is uneconomic or dangerous to approach these relay networks for their adverse circumstance, therefore, recharging or replacing batteries is difficult or impossible. As a result, the lifetime of the relay networks is limited~\cite{Qzhao}. \par

Recently, wireless power transfer and radio-frequency (RF) energy harvesting provide appealing ways to recharge the battery and prolong the lifetime. RF signals carry both information and energy, thus, the idea of simultaneous wireless information and power transfer (SWIPT) is proposed in~\cite{Survey1}and~\cite{Survey2}. Two practical receiving schemes for SWIPT, time switching (TS) and power splitting (PS), are proposed in~\cite{Survey3}. Furthermore, \cite{Survey4}-\cite{Surveyhu} study the dynamic time switching (DTS) scheme and dynamic power splitting (DPS) scheme to improve the quality of information transmission. In addition, the SWIPT in the multi-carrier system \cite{Survey5} and multi-antenna system~\cite{ob} etc., are also discussed. \par

Nasir \it{et al}\upshape~studies the relay network for SWIPT and propose time switching-based relaying (TSR) and power splitting-based relaying (PSR) protocols in \cite{relay}. In \cite{ZhiguoDing}, optimal power allocation at the relay with wireless energy harvesting is studied for multi-user transmission. The SWIPT in two-way relay system is analyzed in \cite{BiaoWang}.
In order to improve the system throughput, adaptive time-switching protocols are also proposed in \cite{DTSR}. Distributed power splitting for SWIPT in multi-user multi-relay networks is proposed in \cite{He}. Since the PSR protocol outperforms the TSR protocol~\cite{relay}, we just focus our attention on the PSR protocol based relay networks. In this paper, we consider a wireless relay network using PSR protocol to harvest energy and adopting amplify-and-forward (AF) scheme to relay signal. The relay has the ability of dynamically adjusting the power splitting ratio of information transmission (IT) and energy harvesting (EH) in order to minimize the outage probability of IT. According to the knowledge of channel state information (CSI) at the relay, optimal dynamic power splitting policies with full CSI and partial CSI are both provided. Through simulations, both proposed policies indeed can improve the outage performances and the policy with full CSI outperforms that with partial CSI.
\section{System Model} \label{sec:2}
\begin{figure}[t]
\centering
\includegraphics[scale=0.55]{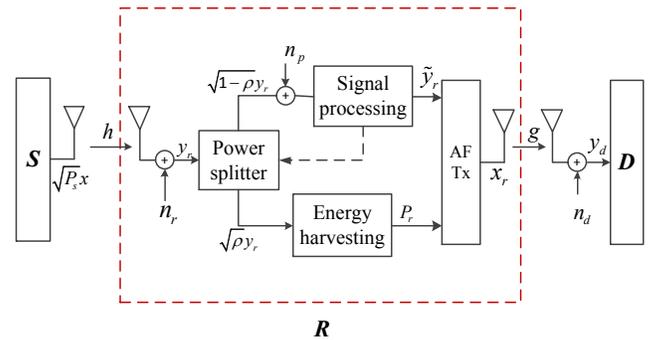}
\caption{System model}
\label{fig_model1}
\end{figure}
In this paper, the AF scheme based wireless relay system with EH is considered, as shown in Fig.\ref{fig_model1}. The source (S) transmits the information to the destination (D), and we assume that the direct link between the source and the destination is not existing. The relay runs the amplify-and-forward (AF) scheme because of its property of low complexity \cite{jing}. All nodes are equipped with one antenna and an storage battery to implement energy charging and discharging.
The energy for the wireless node operating, e.g., circuit running, is far less than that for signal transmitting. We assume the node running is only supported by the initial energy stored in the battery and the initial value can keep the node running in a large enough duration. Accordingly, the relay needs to harvest energy from the received wireless signals for signal forwarding. Each transmission duration $T$ is equally divided into two phases, the first $T/2$ is used for the source to transmit wireless signal to the relay and the other phase is assigned for the relay node to forward information to the destination. Here $h$ denotes the channel gain from the source to the relay and $g$ denotes the channel gain from the relay to the destination. We assume both channels experience the Raleigh fading and keep constant during the two phases~\cite{relay}. As a result, $|h|^2$ and $|g|^2$ follow the exponential distribution with the mean value $\lambda_h$ and $\lambda_g$, respectively.\par

In the first phase, the source transmits the signal $x(t)$ to the relay with power $P_s$. The received RF signal at the relay is
\begin{equation}
y_r(t)=\sqrt{P_s}hx(t)+n_r(t),
\end{equation}
where $n_r(t)$ is the antenna noise and follows the Gaussian distribution with zero mean and variance $\sigma_r^2$. The relay splits the received RF signal into two streams with the power ratio $\rho\in[0,1]$ by the power splitter. After that, $\sqrt{\rho}y_r(t)$ is used for the EH circuit. According to \cite{Survey3}, the harvested energy at the relay is
\begin{equation}
E_h=\varepsilon\rho \left(P_s|h|^2+\sigma_r^2\right)\frac{T}{2},
\end{equation}
where $\varepsilon$ $(0<\varepsilon\le 1)$ denotes the energy converting coefficient of EH circuit. For simplicity, we assume $\varepsilon=1$ in the rest of this paper. The whole harvested energy is used to forward information to the destination by the relay during the second phase \cite{relay}-\cite{He}. Thus, the transmit power of the relay is $P_r=\rho(P_s|h|^2+\sigma_r^2)$. For a practical consideration, $P_r$ should be larger than zero, so that we let $0< \rho\leq 1$. In the meantime, the other RF signal $\sqrt{1-\rho}y_r(t)$ is fed into the signal processing circuit. Then the baseband signal at the relay can be expressed as
\begin{equation}
\tilde y_r=\sqrt{1-\rho}\left(\sqrt{P_s}hx+n_r \right)+n_p,
\end{equation}
where $n_p$ is the additional baseband Gaussian noise with zero mean and variance $\sigma_p^2$ and $x$ and $n_r$ are baseband expressions of $x(t)$ and $n_r(t)$, respectively. If $\rho=1$, it is to say that the relay stops transmitting signal to the destination and harvests all the received energy. Consequently, in this case, there is $\tilde y_r=0$ and the relay closes the information processing circuit. Only if $0<\rho<1$, the AF scheme is activated by the relay. Following the AF scheme~\cite{jing}, the baseband signal transmitted by the relay is
\begin{equation}\label{eq:af}
x_r=\sqrt{P_r}\beta(\rho) \tilde y_r
\end{equation}
where $$\beta(\rho)=\sqrt{\frac{1}{(1-\rho)(P_s|h|^2+\sigma_r^2)+\sigma_p^2}}$$
is the power normalization factor of the AF scheme.\par

In the second phase, $x_r$ is received by the destination through the channel $g$. Hence, the received signal at the destination is
\begin{equation}\begin{split}
y_d &=gx_r+n_d\\
&=\sqrt{\frac{(1-\rho)P_sP_r}{(1-\rho)(P_s|h|^2+\sigma_r^2)+\sigma_p^2}}hgx+\\
&g\sqrt{\frac{P_r}{(1-\rho)(P_s|h|^2+\sigma_r^2)+\sigma_p^2}}(\sqrt{1-\rho}n_r+n_p)+n_d
\end{split}\end{equation}
where $n_d$ is additive Gaussian noise with zero mean and variance $\sigma_d^2$. Then,
the signal-to-noise ratio (SNR) at the destination can be derived as
\begin{equation}
\gamma(\rho)=\frac{P_s|h|^2|g|^2}{|g|^2\sigma_r^2+\frac{|g|^2\sigma_p^2}{1-\rho}+\frac{\sigma_d^2}{P_r\beta(\rho)^2(1-\rho)}}.
\label{gamma}
\end{equation}\par
It is easy to see that $\gamma(\rho)$ is related with the value of $\rho$. In other words, power splitting factor $\rho$ affects the system performances drastically. Therefore, we need to figure out the problem that how to split the signal power so as to achieve the optimal transmission performances. We use the outage probability at the destination to indicate the system performances. Observe (\ref{gamma}), we can see that $\gamma(\rho)$ is also the function of $h$ and $g$. If the relay has the ability to adjusting its $\rho$ according to the channel state, the outage performance can be optimized. Given the available knowledge of CSI at the relay, we consider two cases to study the optimal power splitting.

\section{The Full CSI Case}\label{sec:full}
In this section, we assume that the relay node can obtain full CSI, i.e., $h$ and $g$, before the entire transmission. To obtain the values of $h$ and $g$ at the relay, a RTS (request-to-send)/CTS (clear-to-send) based channel estimation scheme can be employed like~\cite{CTS} and~\cite{zhou}. Before starting the information transmission, the source sends a RTS packet to the relay and the destination. The relay can estimate the channel gain $h$. After receiving the RTS, the destination returns a CTS packet to the source and the relay. Similarly, the relay can also estimate $g$ by itself. We assume the channel estimation is perfect herein. And the effect of channel estimation error is out of the scope of our paper. Through the RTS/CTS mechanism, which is compatible with IEEE 802.11 series standards, the relay can obtain the full CSI before the entire transmission.  Surely, the RTS/CTS mechanism incurs extra overhead and energy consumption for channel estimation. \par

For the relay has the knowledge of $h$ and $g$,  minimizing the outage probability is equivalent to maximize the instantaneous SNR at the destination. Therefore, the optimization problem can be written as
\begin{equation*}
\begin{aligned}
\mathop{\mathrm{Maximize}}_{0 < \rho < 1}\ \ \gamma(\rho)\\
\end{aligned}
\end{equation*}
By (\ref{gamma}), obviously,  $\gamma(\rho)$ is a nonnegative continuous function about $\rho$ and there are $\gamma(0)=0$ and $\gamma(1)=0$. As a result, there must exist the maximum(s) $\{\rho^*\}$ which meet(s) $\frac{\partial \gamma(\rho)}{\partial\rho}|_{\rho=\rho^*}=0$, by the Mean-Value Theorem \cite{MVT}. That is to say the optimal power splitting ratio achieving maximum SNR is one of the roots. In fact, $\gamma(\rho)$ is a concave function on $0<\rho<1$ by calculating its second order derivative. However, the derivations are too complex to be expressed in this paper. Alternatively, we intend to calculate all possible $\rho$ belonging to $$\left\{\rho^*\left|\left.\frac{\partial \gamma(\rho)}{\partial\rho}\right|\right._{\rho=\rho^*}=0, 0< \rho <1\right\}$$
and pick the optimal $\rho^*$ which produces the maximum SNR.\par

Substitute the expression of $P_r$ into (\ref{gamma}), we have
\begin{equation}\begin{split}
&\gamma(\rho)=\\&\frac{P_s|h|^2|g|^2\rho(1-\rho)}{-|g|^2\sigma_r^2\rho^2+(|g|^2(\sigma_r^2+\sigma_p^2)-\sigma_d^2)\rho+\sigma_d^2+\frac{\sigma_p^2\sigma_d^2}{|h|^2P_s+\sigma_r^2}}.
\label{gamma2}
\end{split}\end{equation}
Then the first order derivative of $\gamma(\rho)$ with respect to $\rho$ is derived as (\ref{eq:SNR}).
\newcounter{mytempeqncnt}
\begin{figure*}[!t]
\normalsize
\setcounter{mytempeqncnt}{\value{equation}}
\begin{equation}\label{eq:SNR}
\frac{\partial\gamma(\rho)}{\partial\rho}=\frac{P_s|h|^2|g|^2\left((\sigma_d^2-|g|^2\sigma_p^2)\rho^2-2\left(\sigma_d^2+\frac{\sigma_p^2\sigma_d^2}{|h|^2P_s+\sigma_r^2}\right)\rho+\sigma_d^2+\frac{\sigma_p^2\sigma_d^2}{|h|^2P_s+\sigma_r^2}\right)}{\left(-|g|^2\sigma_r^2\rho^2+(|g|^2\sigma_r^2+|g|^2\sigma_p^2-\sigma_d^2)\rho+\sigma_d^2+\frac{\sigma_p^2\sigma_d^2}{|h|^2P_s+\sigma_r^2}\right)^2}.
\end{equation}
\hrulefill
\vspace*{4pt}
\end{figure*}
To obtain extreme values, we need to solve the equation $\frac{\partial \gamma(\rho)}{\partial\rho}=0$.
Observe (\ref{eq:SNR}), we can see the sign of $\frac{\partial\gamma(\rho)}{\partial\rho}$ corresponds with the numerator term  $f(\rho)=a_1\rho^2+b_1\rho+c_1$, where $$a_1=\sigma_d^2-|g|^2\sigma_p^2,$$ $$b_1=-2\left(\sigma_d^2+\frac{\sigma_p^2\sigma_d^2}{|h|^2P_s+\sigma_r^2}\right),$$ $$c_1=\sigma_d^2+\frac{\sigma_p^2\sigma_d^2}{|h|^2P_s+\sigma_r^2}.$$  Hence we just need to solve $f(\rho)=0$ . If $a_1=0$, then $f(\rho)=0$ becomes a linear equation on $\rho$ and the root is $\rho^*=\frac{1}{2}$. If $a_1>0$, there are two possible roots of $f(\rho)=0$. In this case, as $\frac{-b_1+\sqrt{b_1^2-4a_1c_1}}{2a_1}>1$, we have $\rho^*=\frac{-b_1-\sqrt{b_1^2-4a_1c_1}}{2a_1}$; If $a_1<0$, there is  $\frac{-b_1+\sqrt{b_1^2-4a_1c_1}}{2a_1}<0$, so we have $\rho^*=\frac{-b_1-\sqrt{b_1^2-4a_1c_1}}{2a_1}$. In both cases, $\rho^*$ has the same expression. So the optimal $\rho^*$ which can achieve the maximum SNR is
\begin{equation}\label{fullcsi}
\rho^* = \begin{cases}
\frac{1}{2}, & \text{if }a_1=0;\\
\frac{-b_1-\sqrt{b_1^2-4a_1c_1}}{2a_1}, & \text{else }.
\end{cases}
\end{equation}\par
After obtaining all CSIs based on the RTS/CTS mechanism, the relay can compute the parameters $a_1$, $b_1$ and $c_1$, respectively. Based on the value of $a_1$, the relay adjust its power splitter according to (\ref{fullcsi}) before the entire transmission. After that, the transmission process described in system model section is activated.

\section{The Partial CSI Case}\label{sec:partial}
Since the RTS/CTS mechanism incurs extra overhead for channel estimation and may degrade the transmission efficiency, especially in fast time-varying channel situation. In consideration of the relay harvesting energy from the signal transmitted by the source, the channel coefficient $h$ can be estimated by the relay at the beginning of the first phase. For example, few pilot symbols transmitted by the source can be used to estimate the channel coefficient $h$ by the relay before the entire transmission. Usually, the number of pilot symbols is far smaller than the number of information symbols, so the system cost is very slight. To avoid the signaling exchanging for estimating $g$, we consider the relay only knows the statistic characteristics of $g$, which can be estimated and reported by the destination in a periodic manner~\cite{jing}. As the statistic characteristics of $g$ changes very slowly, the period could be so large that we can ignore the overhead for informing $\lambda_g$ to the relay. Accordingly, the relay knows partial CSI, i.e., $h$ and $\lambda_g$, and can adjust its $\rho$ to achieve the minimum outage probability before the entire transmission.\par

In this case, the optimal policy is to find a $\rho$ to minimize the average outage probability with respect to $g$. Denote the target receiver SNR as $\gamma_0 > 0$, therefore, the optimization problem can be expressed as
\begin{equation}
\begin{aligned}
\label{eq:P2}
\mathop{\mathrm{Minimize}}_{{0<\rho<1}}\ \ &\mathbb{E}_g\left[\Pr(\gamma(\rho)<\gamma_0|h)\right]\\
\end{aligned}
\end{equation}\par
By (6), we have
\begin{equation}
\Pr(\gamma(\rho)<\gamma_0|h)=\Pr\left(|g|^2F(\rho)<\gamma_0\sigma_0^2(\rho)|h\right),
\label{out2}
\end{equation}
where
\begin{subequations}
\begin{align}
&F(\rho)=P_s|h|^2\rho(1-\rho)-\gamma_0\left(-\rho^2\sigma_r^2+\rho\sigma_r^2+\rho\sigma_p^2\right) \label{eq:F},
\\&\sigma_0^2(\rho)=\sigma_d^2(1-\rho)+\frac{\sigma_p^2\sigma_d^2}{|h|^2P_s+\sigma_r^2}.
\label{h0A1b}
\end{align}
\end{subequations}
Note that both $F(\rho)$ and $\sigma_0^2(\rho)$ are independent on $g$. When the relay receives the transmitted signal, $h$ is identified and the relay can only adjust the values of $F(\rho)$ and $\sigma_0^2(\rho)$ by $0 < \rho < 1$. Moreover, on account of $\sigma_0^2(\rho)>0$, there is $\gamma_0\sigma_0^2(\rho) > 0$. To calculate (\ref{eq:P2}), we need to identify the sign of $F(\rho)$. If  $F(\rho) \leq 0$, then $|g|^2F(\rho) < \sigma_0^2(\rho)\gamma_0$ is always true, i.e., $\Pr(|g|^2F(\rho) < \sigma_0^2(\rho)\gamma_0|h)=1$, which means that the destination cannot receive error-free information. To avoid this situation, we should find the optimal ratio in the set ${\Omega}=\{\rho |0<\rho<1,F(\rho)>0 \}$.  Through (\ref{eq:F}), the set $\Omega$ can be written as $\{\rho|0 < \rho < \rho_{\max}(|h|^2)\}$, where $F(\rho_{\max}(|h|^2))=0$ and
\begin{equation}
\label{eq:rho}\rho_{\max}(h)=\frac{P_s|h|^2-\sigma_r^2\gamma_0-\sigma_p^2\gamma_0}{P_s|h|^2-\sigma_r^2\gamma_0}.
\end{equation}
Obviously, if $\rho_{\max}(h) \leq 0$, $\Omega$ becomes an empty set.
By (\ref{eq:rho}), there is a threshold
$H_{0}=\frac{\gamma_0\left(\sigma_r^2+\sigma_p^2\right)}{P_s}$, which is the root of $\rho_{\max}(H_0)=0$, for $h$ in order to make $\Omega$ not empty. If $|h|^2\le H_0$, the transmission outage must be produced no matter how we adjust $\rho$. In other words, there is no room to minimize the outage probability in this case. As a reasonable and practical strategy,  if $|h|^2 \le H_0$ we set $\rho^*=1$ to harvest energy as much as possible. If $|h|^2> H_0$, $\Omega$ is not empty and we need to find out the optimal $\rho$ in $\Omega$. Therefore, we have
\begin{equation}
\begin{aligned}
&\mathbb{E}_g\left[\Pr(\gamma(\rho)<\gamma_0|h)\right]|_{\rho\in\Omega}=\Pr\left(|g|^2< \left.\frac{\sigma_0^2(\rho)\gamma_0}{F(\rho)}\right|h\right)\\
&=\displaystyle{\int_{0}^{\frac{\gamma_0\sigma_0^2(\rho)}{F(\rho)}}\frac{e^{-\frac{x}{\lambda_g}}}{\lambda_g}dx}=1-\exp\left\{-\frac{\gamma_0\sigma_0^2(\rho)}{F(\rho)\lambda_g}\right\}.
\end{aligned}
\end{equation}
Let $W(\rho)=\frac{F(\rho)}{\sigma_0^2(\rho)}$. As minimizing $1-\exp\left\{-\frac{\gamma_0\sigma_0^2(\rho)}{F(\rho)\lambda_g}\right\}$ is equivalent to maximize $W(\rho)$, solve problem (\ref{eq:P2}) equals to
\begin{equation}
\begin{aligned}
\mathop{\mathrm{Maximize}}_{{\rho\in\Omega}}\ \ W(\rho)
\end{aligned}
\end{equation}
The object $W(\rho)\ge0$ is a continuous function of $\rho\in\Omega$. In addition, it is easy to see that $F(0)=F(\rho_{\max})=0$, so that there are $W(0)=0$ and $W(\rho_{\max})=0$. Besides, as $W(\rho)$  is the concave function on $0\leq\rho\leq 1$~\cite{Survey8}, there must exist one $\rho^*$ to maximize $W(\rho)$ over the set $\Omega$. The derivative of $W(\rho)$ with respect to $\rho$ is
\begin{equation}
\frac{\partial W(\rho)}{\partial\rho}=a_2-\frac{c_2}{(\rho-b_2)^2}.
\end{equation}
where $$a_2=\frac{P_s|h|^2-\gamma_0\sigma_r^2}{\sigma_d^2},$$ $$b_2=1+\frac{\sigma_p^2}{|h|^2P_s+\sigma_r^2},$$ $$c_2=b_2\left(\frac{a_2\sigma_p^2}{P_s|h|^2+\sigma_r^2}+\frac{\gamma_0\sigma_p^2}{\sigma_d^2}\right).$$
Let $\frac{\partial W}{\partial\rho}=0$, therefore, we can obtain the roots $\rho=b_2\pm\sqrt{\frac{c_2}{a_2}}$. Meanwhile, as $b_2+\sqrt{\frac{c_2}{a_2}}>b_2>1$, we just choose $\rho^*=b_2-\sqrt{\frac{c_2}{a_2}}$. In summary,
\begin{equation}\label{partialcsi}
\rho^* = \begin{cases}
1, & \text{if}~|h|^2 \leq H_0;\\
b_2-\sqrt{\frac{c_2}{a_2}}, & \text{else }.
\end{cases}
\end{equation}
Through the mechanism introduced at the beginning of this section, the relay can obtain $h$ and $\lambda_g$ so that $\rho^*$ can be calculated by (\ref{partialcsi}). Accordingly, the relay also can adjust its power splitting ratio before the transmission.

\section{Simulation}\label{sec:5}
In this section, we present simulation results to verify the proposed optimal power splitting policies. Simulation parameters are given as: $\sigma_r^2=-20$dBm, $\sigma_p^2=-20$dBm, $\sigma_d^2=-17$dBm ~\cite{relay}. At the source, we set the fixed transmission rate $R=3$ bits/sec/Hz so that the threshold value of SNR at the destination $\gamma_0=2^R-1$.
In order to show the advantage of our proposed dynimic power splitting policies, the fixed power splitting schemes proposed in \cite{relay} with $\rho=0.4$, $\rho=0.6$, and $\rho=0.8$ are also simulated. We simulate the average outage probability of the relay network with above power splitting schemes over $10^{6}$ channel realizations.
\begin{figure}[t]
\centering
\includegraphics[width=9.5cm]{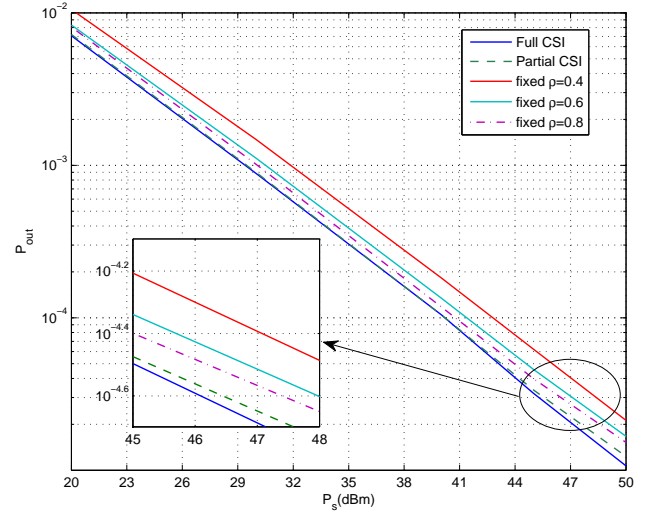}
\caption{Outage probability $P_{out}$ versus transmit power $P_s$. ($\sigma_r^2=-20$dBm, $\sigma_p^2=-20$dBm, $\sigma_d^2=-17$dBm, $\lambda_h=1.5$, $\lambda_g=1.5$).}
\label{P_out_P_s}
\end{figure}

\begin{figure}[t]
\centering
\includegraphics[width=9.5cm]{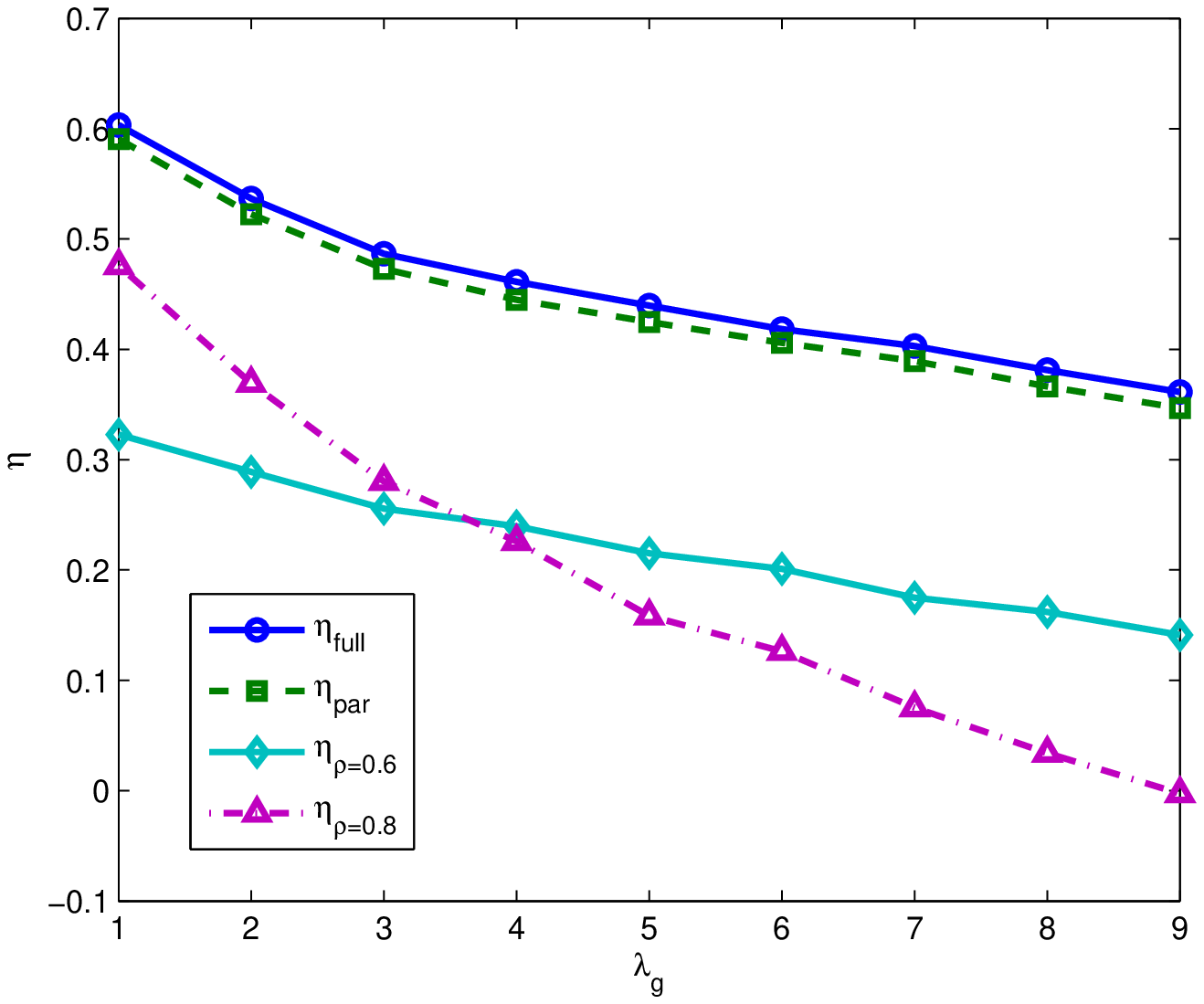}
\caption{Performance gains as $\eta$ versus $\lambda_g$. ($P_s=40$dBm, $\sigma_r^2=-20$dBm, $\sigma_p^2=-20$dBm, $\sigma_d^2=-17$dBm, $\lambda_h=1.5$).}
\label{eta_lambda_g}
\end{figure}

\begin{figure}[t]
\centering
\includegraphics[width=9cm]{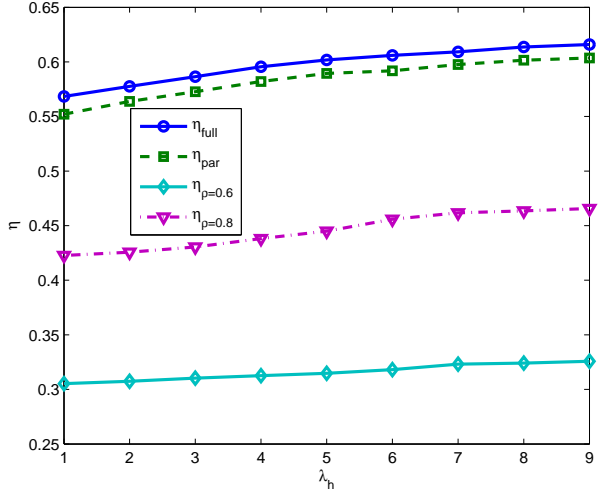}
\caption{Performance gains as $\eta$ versus $\lambda_h$. ($P_s=40$dBm, $\sigma_r^2=-20$dBm, $\sigma_p^2=-20$dBm, $\sigma_d^2=-17$dBm, $\lambda_g=1.5$).}
\label{eta_lambda_h}
\end{figure}

First, we assume the distance from the source to the relay equals to that from the relay to the destination, e.g.,  $\lambda_h=\lambda_g=1.5$.  The outage performances achieved by the proposed power splitting policies are shown in Fig. \ref{P_out_P_s}.
As the transmit power increases, the average outage probability decreases. Both proposed power splitting policies achieve better performances than all the schemes with fixed $\rho$. For example, the full CSI based policy obtain about 1.25 dB gain than the scheme with $\rho=0.8$, 1.7 dB gain than the scheme with $\rho=0.6$ and 2.5 dB gain than the scheme with $\rho=0.4$ if $P_s=50 $dBm. Although the policy with full CSI outperforms the policy with partial CSI, the performance gap is slight, e.g., if $P_s=50$dBm and both average outage probabilities are lower than $10^{-4}$, the gap is only about 0.5 dB. That is to say the policy with partial CSI, which incurs less overhead, approaches the policy with full CSI closely. If the RTS/CTS mechanism is not available, policy with partial CSI is suggested.\par

Second, we investigate the effects of the channel statistical parameters on the gains of the proposed power splitting policies. Treating the policy with $\rho=0.4$ achieving the worse performance as the baseline, we define the performance gain as
\begin{equation*}
\begin{aligned}
&\eta_{full}=-\ln\frac{P_{out}^{full}}{P_{out}^{\rho=0.4}},\ \ \eta_{par}=-\ln\frac{P_{out}^{par}}{P_{out}^{\rho=0.4}},\\
&\eta_{\rho=0.6}=-\ln\frac{P_{out}^{\rho=0.6}}{P_{out}^{\rho=0.4}},\ \ \eta_{\rho=0.8}=-\ln\frac{P_{out}^{\rho=0.8}}{P_{out}^{\rho=0.4}}.
\end{aligned}
\end{equation*} where $\eta_{full}$ is the gain of policy with full CSI, $\eta_{par}$ is the gain of policy with partial CSI, $\eta_{\rho=0.8}$ is the gain of policy with $\rho=0.8$, and $\eta_{\rho=0.6}$ is the gain of policy with $\rho=0.6$.
Fig. \ref{eta_lambda_g} and Fig. \ref{eta_lambda_h} show the performance gains for different $\lambda_g$ and $\lambda_h$. In Fig. \ref{eta_lambda_g}, we can see that as the $\lambda_g$ increases the gains decrease. It means that the proposed dynamic policies are very suitable for the case where the path loss from the relay to the destination is drastic. And both proposed policies have almost the same gains, which are larger than $\eta_{\rho=0.6}$ and $\eta_{\rho=0.8}$. Note that if $\lambda_g \geq4$, there is $\eta_{\rho=0.6} > \eta_{\rho=0.8}$, which means the policy with $\rho=0.6$ achieves better outage performance than the policy with $\rho=0.8$. If $\lambda_g \geq 9$, there is $\eta_{\rho=0.8}<0$. That is to say the policy with $\rho=0.8$ has a worse outage performance than the policy with $\rho=0.4$.  In Fig. \ref{eta_lambda_h}, the performance gains increases as $\lambda_h$ increases. The reason is that the more energy the relay obtains the more room the proposed policies have for transmission optimization. Similarly, $\eta_{par}$ approaches $\eta_{full}$ closely. Both proposed policies outperform the policies with fixed $\rho$. From both figures, we can see that the channel parameters affect the performances of the proposed power splitting policies. The policy with full CSI achieves the best performances and the latter approaches the former closely all the time.

\section{Conclusion}\label{sec:6}
In this paper, dynimic power splitting policies with full CSI and partial CSI for rely networks with wireless energy harvesting are proposed to minimize the system outage probability.
Through simulations, it is found that both proposed policies outperform the policy with fixed power splitting ratio. Although the policy with full CSI can achieve the best performance, extra system overhead is also incurred. Bring in negligible cost, the policy with partial CSI approaches the best performance closely.

\section*{Acknowledgment}
This work is supported by the Research Fund of National Mobile Communications Research Laboratory, Southeast University (No. 2011D14).

\end{document}